\documentclass{aa}

\usepackage{graphicx}
\usepackage{txfonts}
\begin{document} 

   \title{The ESO-VLT MIKiS survey reloaded: the internal kinematics of the core of M75\thanks{Based on observations collected at the European Southern Observatory, Cerro Paranal (Chile), under Large Programme 106.21N5 (PI: Ferraro).}}

   \subtitle{}

   \author{Silvia Leanza\inst{1} \inst{2} \and
          Cristina Pallanca\inst{1} \inst{2} \and
           Francesco R. Ferraro \inst{1} \inst{2} \and
          Barbara Lanzoni \inst{1} \inst{2} \and
          Enrico Vesperini \inst{3} \and
          Mario Cadelano\inst{1} \inst{2} \and          
          Livia Origlia \inst{2} \and
          Cristiano Fanelli \inst{2} \and
          Emanuele Dalessandro \inst{2} \and
          Elena Valenti \inst{4} \inst{5}
          }

   \institute{Dipartimento di Fisica e Astronomia, Universit\`a di Bologna, Via    Gobetti 93/2 I-40129 Bologna, Italy \and
      INAF-Osservatorio di Astrofisica e Scienze dello Spazio di Bologna, Via Gobetti 93/3 I-40129 Bologna, Italy \and    
      Department of Astronomy, Indiana University, Bloomington, IN, 47401, USA \and
      European Southern Observatory, Karl-Schwarzschild-Strasse 2, 85748 Garching bei Munchen, Germany \and
      Excellence Cluster ORIGINS, Boltzmann-Strasse 2, D-85748 Garching Bei Munchen, Germany}

   \date{}

 
  \abstract{We present the results of a study aimed at characterizing
the kinematics of the inner regions of the halo globular cluster M75 (NGC 6864) based on data acquired as part of the ESO-VLT Multi-Instrument Kinematic Survey (MIKiS) of Galactic globular clusters.
Our analysis includes the first determination of the line-of-sight velocity dispersion profile in the core region of M75.  By using
MUSE/NFM observations, we obtained a sample of $\sim 1900$ radial
velocity measurements from individual stars located within $16\arcsec$
(corresponding to about $r <  3 r_c$ where $r_c$ is the estimated core radius of the system)
from the cluster center.  After an appropriate selection of the most
accurate velocity measures, we determined the innermost portion of the
velocity dispersion profile, finding that it is characterized by a
constant behavior and a central velocity dispersion of $\sigma_0\sim
9$ km s$^{-1}$.
The simultaneous King model fitting to the projected velocity
dispersion and density profiles allowed us to check and update
previous determinations of the main structural parameters of the
system.  We also detected a mild hint of rotation in the central $\sim
7\arcsec$ from the center, with an amplitude of just $\sim 1.0$ km
s$^{-1}$ and a position angle of the rotation axis of PA$_0 =
174\degr$. Intriguingly, the position angle is consistent with that
previously quoted for the suspected rotation signal in the outer
region of the cluster.  Taking advantage of the high quality of the
photometric catalog used for the analysis of the MUSE spectra, we also
provide updated estimates of the cluster distance, age, and reddening.}
   

  \maketitle

%

\section{Introduction}
\label{sec:intro}
This paper is part of a long-term project ({\it Cosmic-Lab}) aimed at
using star clusters as
natural cosmic laboratories to study stellar evolution, dynamical
evolution, and how the properties of the hosted stellar populations
evolve with time in high-density environments.  Globular Clusters (GCs) are the ideal
stellar systems for this kind of investigation since the continuous
gravitational interactions among stars typically occur on a timescale
shorter than their age. Hence the effects of this internal evolution
are expected to leave some signatures on their stellar populations,
e.g., via the formation of exotic species like interacting binaries,
blue straggler stars (BSSs) and millisecond-pulsars that cannot be
explained by standard stellar evolution (see, e.g., \citealp{bailyn95,
  pooley+03, ransom+05, ferraro+92, ferraro+03, ferraro+18a}).

The approach proposed by Cosmic-Lab foresees a comprehensive study of
the global properties of each investigated stellar system, including
the photometric and chemical characterization of both the normal and
the exotic stellar sub-populations hosted in it. The adopted approach
can be schematically summarized as follows: (1) the detailed
photometric and spectroscopic characterization of the ``canonical''
sub-populations hosted in the system
\citep[see examples in][]{valenti+10, dalessandro+13a, dalessandro+14, dalessandro+16, dalessandro+22,
  saracino+15, saracino+19, ferraro+97, ferraro+09a,
  ferraro+16,ferraro+21, pallanca+21, cadelano+23,
  deras+24, origlia+97,origlia+02,origlia+03,origlia+11,origlia+13,
  origlia+19, massari+14,crociati+23}; (2) the determination of the
cluster structural parameters via projected density profiles derived
from star counts even for the innermost regions of high-density
stellar systems \citep[see, ][]{ibata+09,lanzoni+07,
  lanzoni+10, lanzoni+19, miocchi+13, cadelano+23, deras+23}; (3) the
characterization of any hosted exotic stellar population \citep[see, e.g.,][]{paresce+92, ferraro+99, ferraro+01,ferraro+03, ferraro+09b,
  ferraro+23a, ferraro+23b, dalessandro+08, dalessandro+13b, cadelano+18, cadelano+20a,
  pallanca+10, pallanca+13, pallanca+14, pallanca+17, billi+23}; (4) the full kinematic characterization of
the detected populations via the construction of velocity dispersion
and rotation profiles from line-of-sight velocities and proper motions
of individual stars \citep[see examples in][]{ferraro+18b, lanzoni+13,
  lanzoni+18a, lanzoni+18b, libralato+18, raso+20, dalessandro+21,
  leanza+22, leanza+23, pallanca+23}.

In this respect the ESO-VLT Multi-Instrument Kinematic Survey 
\citep[MIKiS;][]{ferraro+18b,ferraro+18c} has been specifically designed to
characterize the kinematical properties of a representative sample of
Galactic GCs at different dynamical evolutionary stages, using
different spectrographs currently available at ESO. In the context of
this survey, the detailed investigation of the almost unexplored
innermost cluster regions has been carried out in an ongoing Large
Program (106.21N5, PI: Ferraro) that exploits the remarkable
performance of the adaptive optics (AO) assisted integral field
spectrograph MUSE. 
 
As part of this project, here we discuss the internal kinematics of
the core region of the Galactic GC M75 (NGC 6864). This is a massive
system (with $V-$band absolute magnitude $M_V = -8.57$) of
intermediate/high metallicity ([Fe/H]$=-1.29$ dex; \citealt{carretta+09}),
and relatively high central density ($\mathrm{log}(\rho_0/$M$_\odot$ pc$^{-3})=4.9$; \citealt{pryor+93}). It is located in the Halo of
the Galaxy at a distance $d\sim 20$ kpc from the Sun, under the
assumption of an apparent distance modulus $(m-M)_V=17.09$ and a
color excess $E(B-V)=0.16$ \citep[][2010 edition]{harris+96}.  The
shape of the BSS radial distribution suggested that this is a
dynamically evolved cluster \citep[see][]{contreras+12, ferraro+12},
and this result was later confirmed by the analysis of the BSS central
sedimentation level \citep{lanzoni+16}: indeed, M75 is the second most
dynamically evolved cluster among the approximately 60 stellar systems
investigated so far with this method in the Galaxy \citep{ferraro+18c,
  ferraro+20, ferraro+23a} and in the Magellanic Clouds
\citep{ferraro+19, dresbach+22}.
\section{Observations and data reduction}
\label{sec:obs}
\begin{figure}[ht!]
\centering
\includegraphics[width=0.5\textwidth]{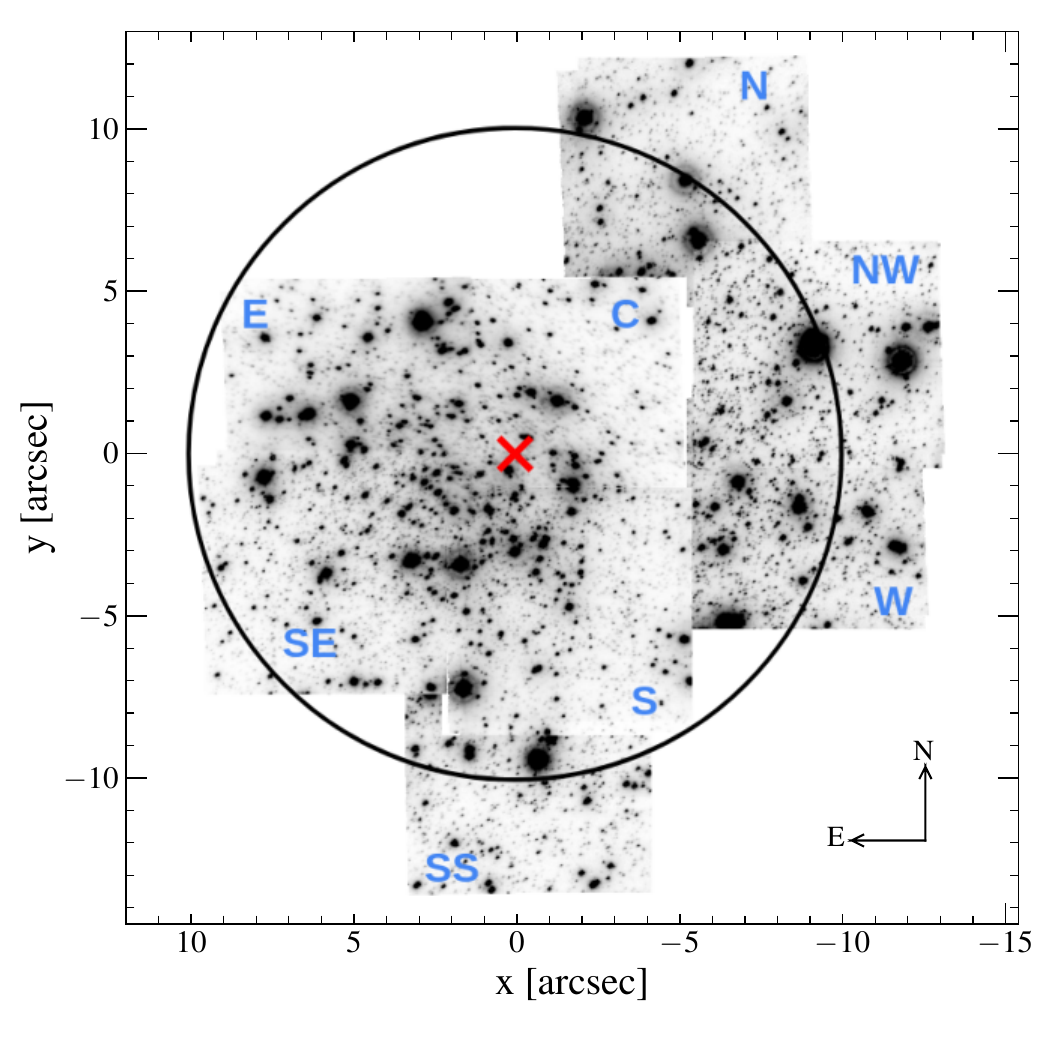}
\centering
\caption{Reconstructed $I$-band images of the MUSE/NFM pointings
  acquired in the core of M75. The circle is centered on the cluster
  center (red cross, from \citealp{contreras+12}) and has a radius of
  $10\arcsec$.}
\label{fig:mosaico}
\end{figure}

To explore the line-of-sight internal kinematics of the innermost
regions of M75, we
acquired spectra of resolved stars with the AO-assisted integral-field spectrograph MUSE \citep{bacon+10} installed
at the Yepun (VLT-UT4) telescope at the ESO Paranal Observatory.  MUSE
is constituted by a modular structure of 24 identical Integral Field
Units (IFUs), and it is available in two configurations: Wide Field
Mode (WFM) and Narrow Field Mode (NFM).  We adopted the MUSE/NFM
configuration, which provides the highest spatial resolution 
(with a spatial sampling of $0.025\arcsec$/pixel) over a field of view
of $7.5\arcsec\times 7.5\arcsec$. The NFM configuration is equipped
with the GALACSI-AO module \citep{Arsenault+08, Strobele+12} and
additionally takes advantage of the Laser Tomography AO correction,
which is not available in the MUSE/WFM configuration.  MUSE/NFM
samples the wavelength range from $4750$ \AA\ to $9350$ \AA, with a
resolving power R $\sim3000$ at $\lambda\sim8700$ \AA.  The data set
was acquired as part of the ESO Large Program ID: 106.21N5.003 (PI:
Ferraro), and consists of a mosaic of eight MUSE/NFM pointings in the
core region of the cluster, as shown in Fig. \ref{fig:mosaico}.  A
series of three 750 s long exposures were secured for each pointing
(see Table \ref{tab:data}), with an average DIMM seeing during the
observations of $0.7\arcsec$.  A rotation offset of $90\degr$ and a
small dithering pattern were set between consecutive exposures, in
order to remove possible systematic effects and resolution differences
between the individual spectrographs.  The data reduction was
performed by using the standard MUSE pipeline \citep{Weilbacher+20}
through the EsoReflex environment \citep{Freudling+13}.  The pipeline
first performs bias subtraction, flat field correction, and wavelength
calibration for each IFU, then applies the sky subtraction, performs
the flux and astrometric calibration, and corrects all data for the
heliocentric velocity. Next, a datacube is produced for each exposure
by combining the pre-processed data from the 24 IFUs. As the last step,
the pipeline combines the datacubes of the multiple exposures of each
pointing into a final datacube, taking into account the offsets and
rotations between the different exposures.

The mosaic of the reconstructed $I$-band images from the stacking of
MUSE datacubes is shown in Fig. \ref{fig:mosaico}. Each pointing is
labeled with a name according to its position with respect to the
cluster center: C, E, N, NW, S, SS, SE, and W
stand for central, eastern, northern, northwestern, southern,
south-southern, southeastern, and western pointing, respectively. As
can be appreciated, the mosaic provides a nice and almost complete
sampling of the innermost $\sim 7\arcsec$ radial region of the
cluster.

\begin{table}
\caption{MUSE/NFM data sets for M75}
\centering
\setlength{\tabcolsep}{14pt} 
\renewcommand{\arraystretch}{1.2} 
\begin{tabular}{clccc}
\hline\hline
Name & & Date & N$_{\rm exp}$ & $t_{\rm exp}$ [s] \\
\hline
C  & & 2021-08-17  & 3 & 750\\
E  & & 2021-08-17 & 3 & 750\\
S  & & 2021-08-17 & 3 & 750\\
N  & & 2021-08-18 & 3 & 750\\
W  & & 2021-09-19 & 3 & 750\\
SS & & 2021-09-19 & 1 & 750\\
NW & & 2022-06-30 & 3 & 750\\
SE & & 2022-07-03 & 3 & 750\\
\hline
\end{tabular}
\tablefoot{For each MUSE/NFM pointing, the table lists the name, execution date, number of exposures (N$_{\rm exp}$) and exposure time of each frame ($t_{\rm exp}$, in seconds).}
\label{tab:data}
\end{table}


\section{Analysis}
\label{sec:analysis}
For the extraction of the MUSE spectra we used the software PampelMuse
\citep{kamann+13}, which is designed to extract deblended source
spectra of individual stars from observations of crowded stellar
regions with integral field spectroscopy, by performing a
wavelength-dependent point spread function (PSF) fitting.

\subsection{The photometric catalog}
PampelMuse needs in input the spectroscopic datacube and a photometric
reference catalog with the coordinates of the stars in the datacube
field of view, and their magnitudes.  We noticed that the photometric
catalog presented in \citet{contreras+12} is not appropriate for this
purpose since many bright stars are lacking because of saturation.
Therefore, 
 we decided to build a new catalog by analyzing the archive HST/WFC3
 images secured by using the F438W and F555W filters (proposal ID:
 11628, PI: Noyola).
 
A total of six images have been analyzed: three images in the F438W
filter with an exposure time of 420 s, and three in the F555W with an
exposure time of 100 s.  The photometric analysis has been performed
using the software DAOPHOT \citep{Stetson+87}.  Briefly, in a first
step the PSF in each image is modeled by using a large sub-sample
($\sim 200$) of bright, isolated, and well-distributed stars. Then, a
star search is performed in each image with a $5\sigma$ threshold
above the background level, and the PSF model previously found is
fitted to all of these identified sources.  Subsequently, we created
as reference a master list including all the sources measured in at
least half of the images acquired with the F555W filter. Using this
master list we run the DAOPHOT/ALLFRAME package \citep{Stetson+94} to
force the fit of the PSF model to the location of these sources in all
the other images.  For each of the identified stars, the magnitude
values measured in the different filters were combined using DAOMATCH
and DAOMASTER.  The final catalog includes the instrumental
coordinates, the mean magnitude in each filter, and the photometric
errors for the detected sources.  The instrumental magnitudes were
then calibrated to the VEGAMAG photometric system by using the
appropriate aperture corrections, zero-points and the procedure
reported on the HST WFC3
website\footnote{\url{https://www.stsci.edu/hst/instrumentation/wfc3/data-analysis/photometric-calibration/uvis-photometric-calibration}}.
The instrumental positions were corrected for the effect of geometric
distortions within the field of view, by applying the correction
coefficients quoted in \citet{bellini+11}.  Finally, the
distortion-corrected positions have been placed onto the absolute
coordinate system ($\alpha$, $\delta$) by cross-correlation with the
\textit{Gaia} DR3 catalog \citep{gaia+23}.

Even in this catalog, however, a few bright stars ($\sim10$) are
saturated. To include them in the analysis, we thus performed a
similar photometric analysis on two 2D MUSE images obtained by
stacking the wavelength slices at 4840-4860 \AA\ and 6060-6080 \AA\ of
each datacube.
For each MUSE pointing we obtained a catalog with the instrumental
positions and magnitudes of the detected stars. Then, the
cross-correlation with the HST/WFC3 catalog discussed above allowed us
to assign absolute coordinates and calibrated magnitudes to every MUSE
star.  We emphasize that we used the MUSE photometry only to recover
the magnitude of the few bright stars that were saturated in the
HST/WFC3 catalog. As the final reference catalog for PampelMuse, we
thus adopted the HST/WFC3 catalog (which guarantees high astrometric
precision and allows us to properly resolve stars even in the most
crowded regions), with the addition of the MUSE measurements obtained
for the few saturated sources.

Besides the reference catalog and the datacube, PampelMuse also
requires in input an analytical PSF model, which is necessary for the
source deblending.  We selected the MAOPPY function \citep{fetick+19}
as PSF model, which accurately reproduces the typical double-component
(core and halo) shape of the AO-corrected PSF in the MUSE/NFM
observations \citep[see][]{gottgens+21}.
Once the inputs are set, for each slice of the MUSE datacube,
PampelMuse performs a PSF fitting using a sub-sample of stars,
obtaining the wavelength dependencies of the PSF parameters. At the
same time, the software fits, for each slice, also the coordinate
transformation from the reference catalog to the data.  Then, these
wavelength-dependent quantities are used to perform the deblending of
all the sources present in the MUSE field of view, and extract the
spectra.
\begin{figure*}[ht!]
\centering
\includegraphics[width=0.8\textwidth]{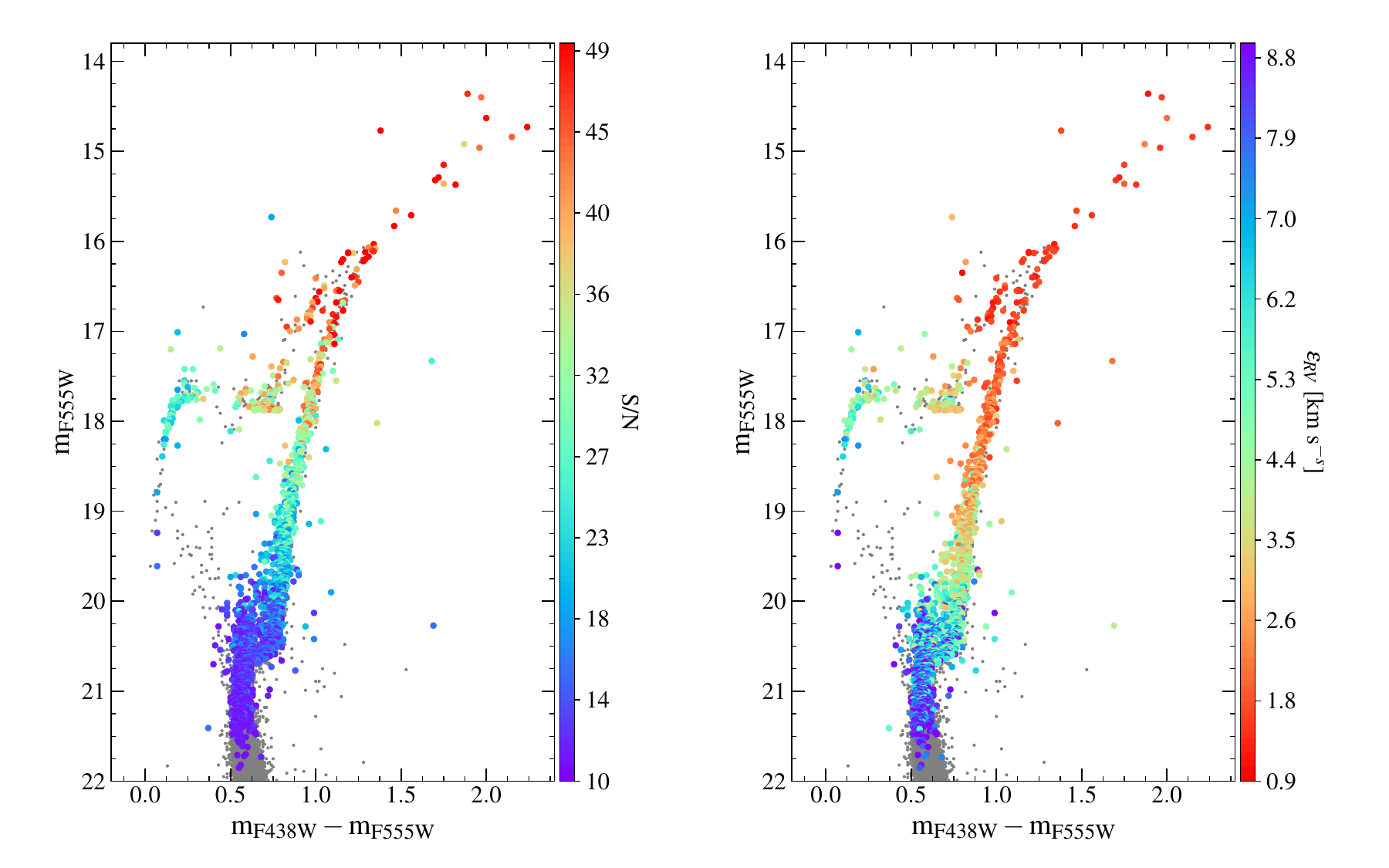}
\centering
\caption{CMD of M75 (gray dots) obtained from the HST/WFC3 photometric
  catalog described in Section \ref{sec:analysis}. The large colored
  circles show the spectroscopic targets of the MUSE catalog (see
  Section \ref{sec:catalog}), while the color scale in the left and
  right panels represents the S/N of the spectra, and the RV
  uncertainty, respectively.}
\label{fig:cmd}
\end{figure*}

\subsection{The measure of radial velocities}
To measure the radial velocity (RV) of the surveyed stars, we applied
the same procedure described in \citet[][see also
  \citealp{pallanca+23}]{leanza+23}, which uses the Doppler shifts of
the Calcium Triplet lines ($8450-8750$ \AA).  The procedure needs a
set of suitable synthetic spectra to be used as reference. We computed
a library of synthetic spectra using the SYNTHE code
(\citealt{sbordone+04} and \citealt{kurucz+05}), adopting an
$\alpha-$enhanced chemical mixture ([$\alpha$/Fe]$=0.4$ dex), the
cluster metallicity [Fe/H]$=-1.29$ dex \citep{carretta+09}, and a set
of atmospheric parameters (effective temperature and gravity)
appropriate for the evolutionary stage of the targets, according to
their position in the CMD.  The spectra have been produced in the same
wavelength range sampled by the observations, and convolved with a
Gaussian profile to reproduce also the MUSE spectral resolution.
Firstly, the observed spectra are normalized to the continuum, which
is computed through a spline fitting of the spectrum in a proper
wavelength range.  The procedure compares the normalized observed
spectra with each synthetic spectrum of the library, which is shifted
in RV by steps of 0.1 km s$^{-1}$ in an adequate range of RV. For each
velocity shift, and each template, the residuals between the observed
spectrum and the synthetic one are computed.  Then, for each target,
the procedure finds the minimum standard deviation among all the
residuals. This value is associated to a specific template and RV
shift, therefore, the procedure derives, as a result, the best-fit
synthetic spectrum (hence, an estimate of temperature and gravity),
and the RV of the target.  The procedure computes also the signal to
noise ratio (S/N) of the spectra as the ratio between the average of
the counts and their standard deviation in the spectral range $8000 -
9000$ \AA.  The S/N of the targets is shown by the color scale in the
left panel of Fig. \ref{fig:cmd}.

We estimated the RV uncertainties ($\epsilon_{\rm RV}$) using Monte
Carlo simulations. We generated $\sim 9000$ simulated spectra with S/N
between 10 and 90, and applied to these spectra the same procedure
used for the observed ones, obtaining a relation to derive the RV
uncertainties \citep[for more details see][]{leanza+23}.  The errors
obtained are of the order of 2 km s$^{-1}$ for the brightest stars,
and increase to $\sim 8$ km s$^{-1}$ for the faintest stars, as shown
by the color scale of the CMD in the right panel of Figure
\ref{fig:cmd}.
\subsection{Final MUSE catalog}
\label{sec:catalog}
In order to obtain a consistent catalog, we checked if the RV
measurements were homogeneous among the different MUSE pointings. To
this aim, exploiting the several overlapping fields in our data set,
we compared the RV values measured for the stars in common between two
adjacent overlapping regions, and we found offsets ranging from 2 to 4
km s$^{-1}$.  These offsets in the RV measurements might be introduced
by variations in the wavelength calibration among the different
pointings, since the MUSE observations were taken during different
nights (in some cases in different months).  To check the accuracy of
the wavelength calibration and correct the offsets, we measured the RV
from the Fraunhofer A telluric absorption bands ($7570-7680$ \AA).  We
estimated the telluric RV for a sub-sample of targets with S/N $> 20$
by using a procedure analogous to that adopted for the determination
of the star RVs.  In this case, a single template spectrum was adopted
for each night. We generated the appropriate synthetics for the
telluric absorption spectra by using the TAPAS
tool\footnote{\url{https://tapas.aeris-data.fr/en/home/}}
\citep{Bertaux+14}, setting the MUSE spectral range and resolution.
For each pointing, we determined the average telluric RV, finding
velocity zero-points from $0.4$ km s$^{-1}$ up to a maximum of $3.3$
km s$^{-1}$.
We then
subtracted the value of each pointing from all the RV measurements of
the stars in that field, in order to align the different pointings.
The comparison of the RV values in two overlapping pointings after the
correction finally showed a good agreement within the errors.

\begin{figure}[ht!]
\centering
\includegraphics[width=0.3\textwidth]{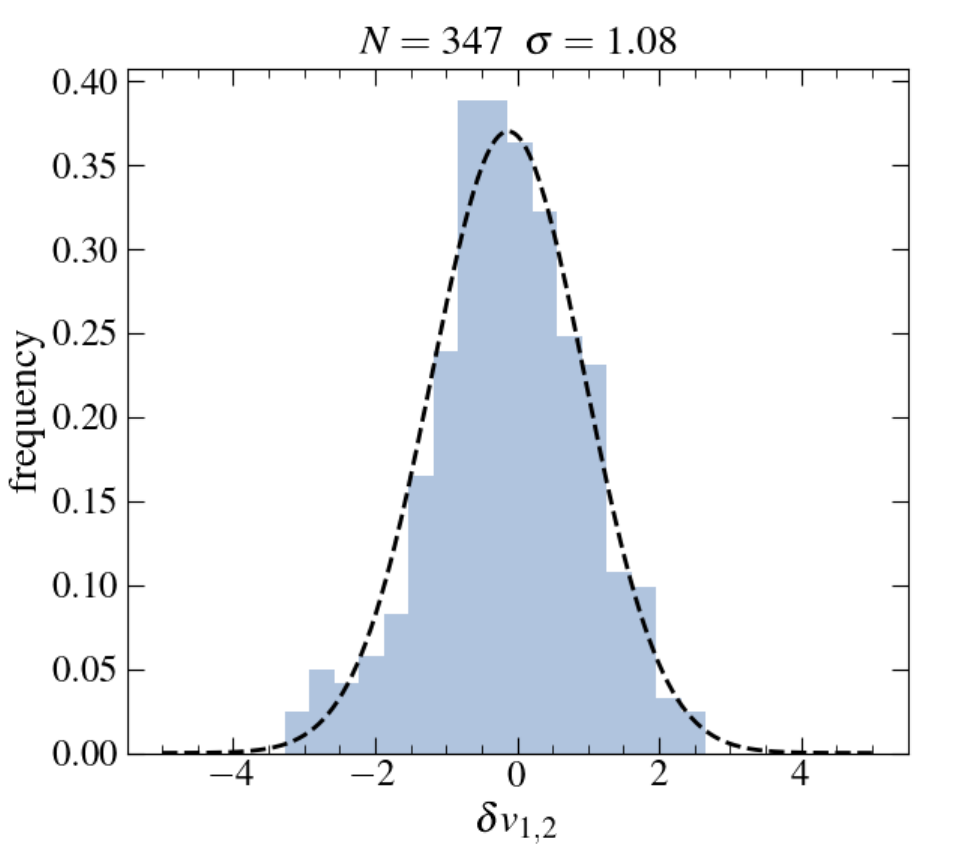}
\centering
\caption{ Histogram of the values of $\delta v_{1,2}$
  obtained from Equation (\ref{eq:errors}) for the stars with multiple
  exposures. The black dashed line represents the Gaussian fit to the
  histogram. Its standard deviation and the number of stellar pairs
  used are labeled at the top of the panel.}
\label{fig:offset}
\end{figure}

Moreover, we used the velocity measures ($v_1$ and $v_2$) of the
targets in common between multiple pointings and their associated
errors ($\epsilon_1$ and $\epsilon_2$) to check whether the estimated
RV uncertainties are reliable. If the quantity
\begin{equation}
\centering
 \delta v_{1,2} = \frac{v_1 - v_2} {\sqrt{\epsilon_1^2 +
     \epsilon_2^2}},
\label{eq:errors}
\end{equation}
results in a normal distribution with unit standard deviation, the
uncertainties are properly estimated \citep[see also][]{kamann+16}.
The resulting distribution is plotted in Fig. \ref{fig:offset} and shows that our uncertainties have been properly
estimated.

\begin{figure}[ht!]
\centering
\includegraphics[width=0.5\textwidth]{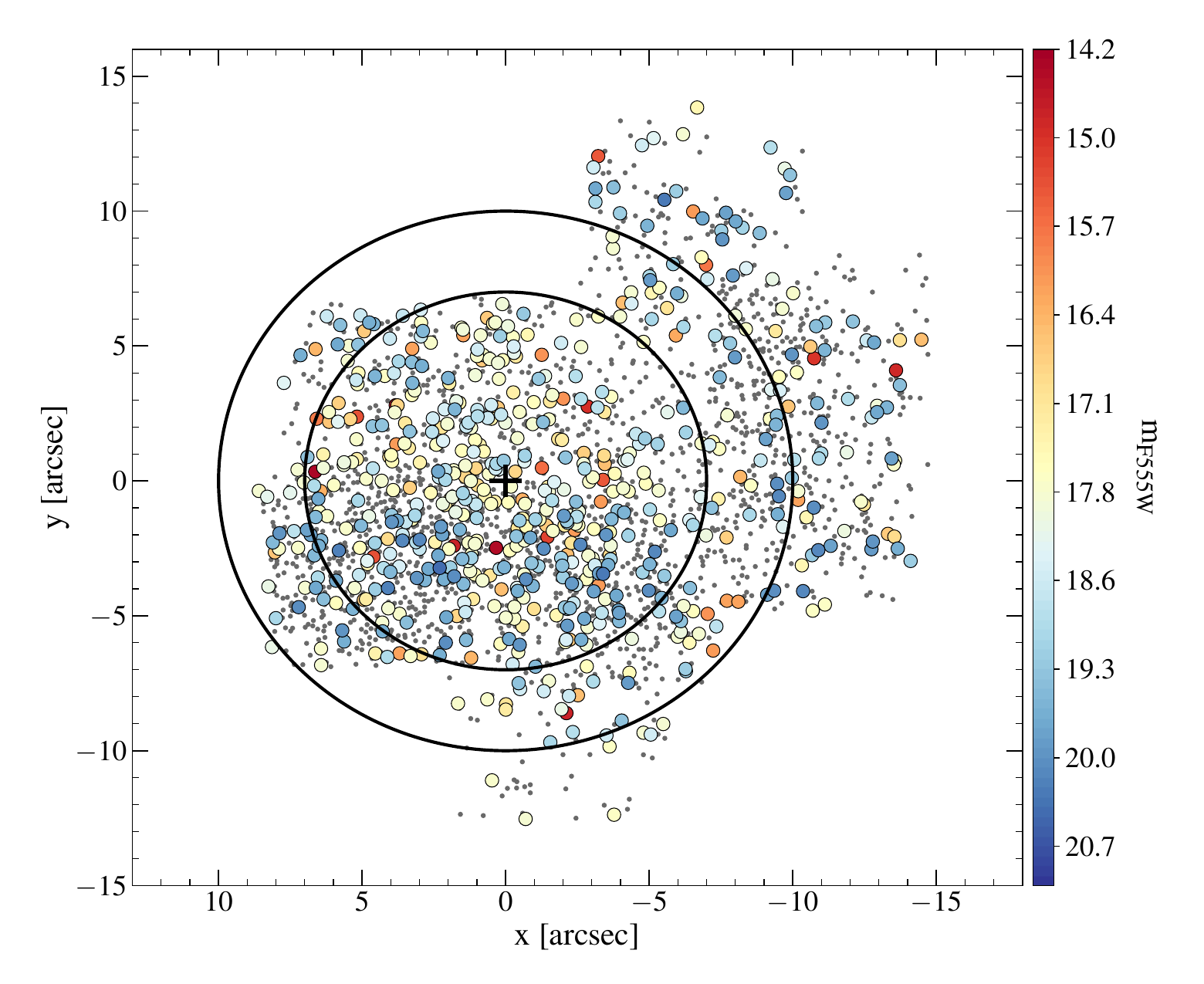}
\centering
\caption{Position of the MUSE targets in the plane of the sky, with
  respect to the cluster center \citep[black cross,][]{contreras+12}.  The
  large colored circles mark the sample of stars that remain after the
  quality selection (S/N $>20$ and RV error $<5$ km s$^{-1}$; see
  Section \ref{sec:results}) and that are used for the kinematic
  analysis, while the gray dots indicate the rejected targets. The
  color scale represents the m$_{\rm F555W}$ magnitude and the two
  circles are centered in the cluster center and have radii of
  7$\arcsec$ and 10$\arcsec$. }
\label{fig:map}
\end{figure}

To construct the final catalog, only stars with S/N $>10$ have been
considered, and for the targets with multiple exposures, the RV value
was determined as the weighted mean of all the available measures, by
using the individual errors as weights.  The resulting MUSE catalog
consists of 1872 RV measurements of individual stars.
The CMD position of these stars is highlighted in Fig. \ref{fig:cmd}
with large circles color-coded according to the S/N values and the RV
uncertainties, in the left and right panels, respectively.  Figure
\ref{fig:map} shows the position of all the MUSE targets (gray
dots and colored circles) in the plane of the sky with respect to the
cluster center, while their RVs as a function of the distance from the
center and their RV distribution are plotted in Fig. \ref{fig:vsys}
(left panel and empty histogram in the right panel, respectively).
 
\section{Results}
\label{sec:results}

For our kinematic study, we have selected only the (630) stars with the best RV measurements\footnote{The sample of bona fide RV measures with the
corresponding errors is publicly available at:
\url{http://www.cosmic-lab.eu/Cosmic-Lab/MIKiS_Survey.html}}  with S/N $>20$ and RV error $<5$ km s$^{-1}$.
The position in the plane of the sky of the selected sample is shown
in Fig. \ref{fig:map} with large colored circles.


\subsection{Systemic velocity}
\label{sec:vsys}
\begin{figure*}[ht!]
\centering
\includegraphics[width=0.85\textwidth]{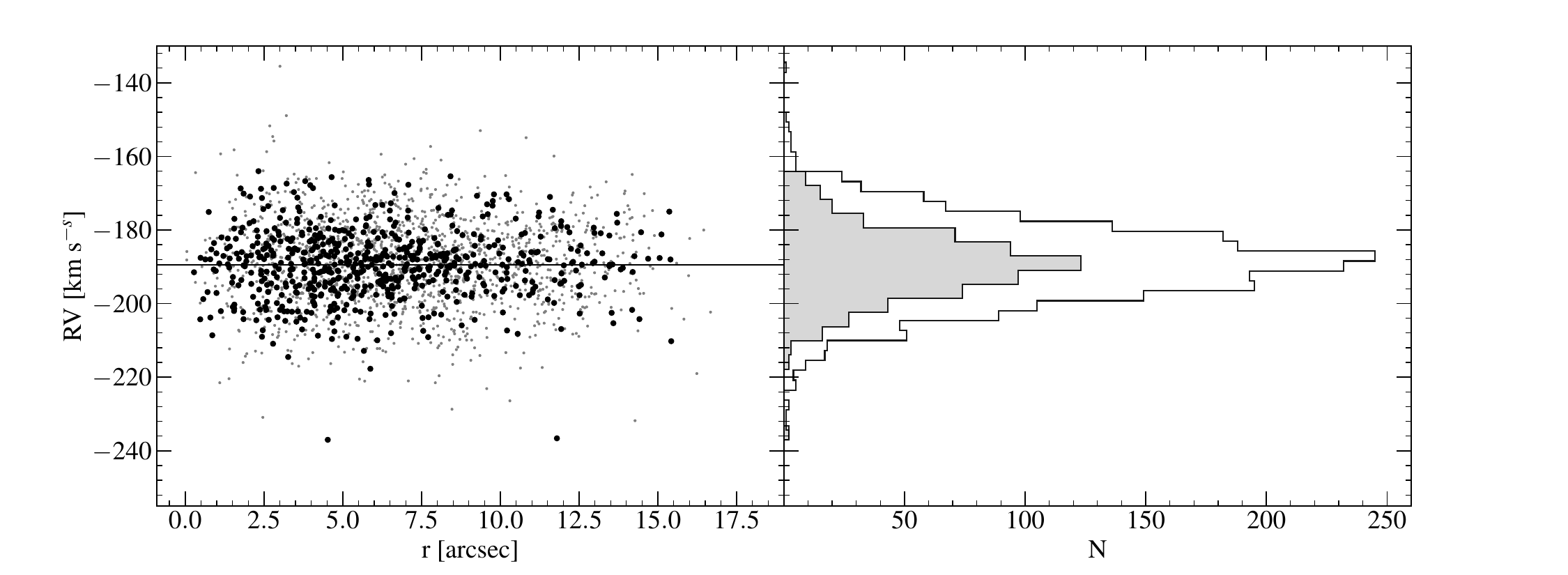}
\centering
\caption{{\it Left panel:} RVs of the final MUSE catalog (see Section
  \ref{sec:catalog}) plotted as a function of the distance from the
  cluster center. The black dots highlight the sample of the well-measured
  stars (i.e., those with S/N $>20$ and RV error $<5$ km
  s$^{-1}$) that is used for the kinematic analysis.  {\it Right panel:}
  the empty histogram shows the number distribution of the entire RV
  sample, while the gray histogram corresponds to the sub-sample of
  bona fide stars (black dots in the left panel).}
\label{fig:vsys}
\end{figure*}

The black dots and the gray shaded histogram in Fig. \ref{fig:vsys}
show the RV distribution of the well-measured stars. As can be
appreciated, the contamination from Galactic field sources is
essentially null for this field of view.
To properly determine the systemic
velocity ($V_{\rm sys}$) of M75 we made the conservative choice of
applying a $2\sigma$-clipping algorithm to the selected targets, thus
removing possible outliers.
Assuming a Gaussian distribution of the velocities, we estimated
$V_{\rm \rm sys}$ and its uncertainty by using a Maximum-Likelihood
algorithm \citep{Walker+06}. The resulting value is $V_{\rm sys} =
-189.5 \pm 0.3$ km s$^{-1}$.  This result is in full agreement with
the values quoted in \citet[][$-188.6 \pm 0.9$ km
  s$^{-1}$]{Baumgardt+18} and \citet[][$-189.3 \pm 3.6$ km
  s$^{-1}$]{harris+96}, while it is consistent within $\sim 2\sigma$
with that derived in \citet[][$-186.2 \pm 1.5$ km s$^{-1}$]{koch+18}.
\subsection{Systemic rotation}
\label{sec:rot}
The internal kinematics of M75 has been little studied so far.  In a
previous work, \citet{koch+18} found an indication of internal
rotation in the outer region of the cluster. However, their analysis
is based on a statistically poor sample (only 32 stars) of RV
measurements.  Taking advantage of our large data set, here we
investigated the possible presence of ordered rotation in the central
regions of the system.

For the analysis of the cluster's internal rotation, we have adopted a method 
often used in the literature (\citealp[see, e.g.,][and references therein;]{leanza+23,pallanca+23} \citealp[see also][for a full description of the method]{cote+95,bellazzini+12}).
Since the proper application of the procedure
requires a uniform distribution of the RV measures in the plane of the
sky, we limited the analysis to the sub-sample of (403) stars located within
$7\arcsec$ 
(corresponding to $r<1.4 r_c$, where $r_c = 4.9 \arcsec$ is the cluster core radius estimated in this work, see Section \ref{sec:discussion} and Fig. \ref{fig:vdp})
from the center, thus avoiding some under-sampled regions
(see Fig. \ref{fig:map}).

\begin{figure*}[ht!]
\centering
\includegraphics[width=0.95\textwidth]{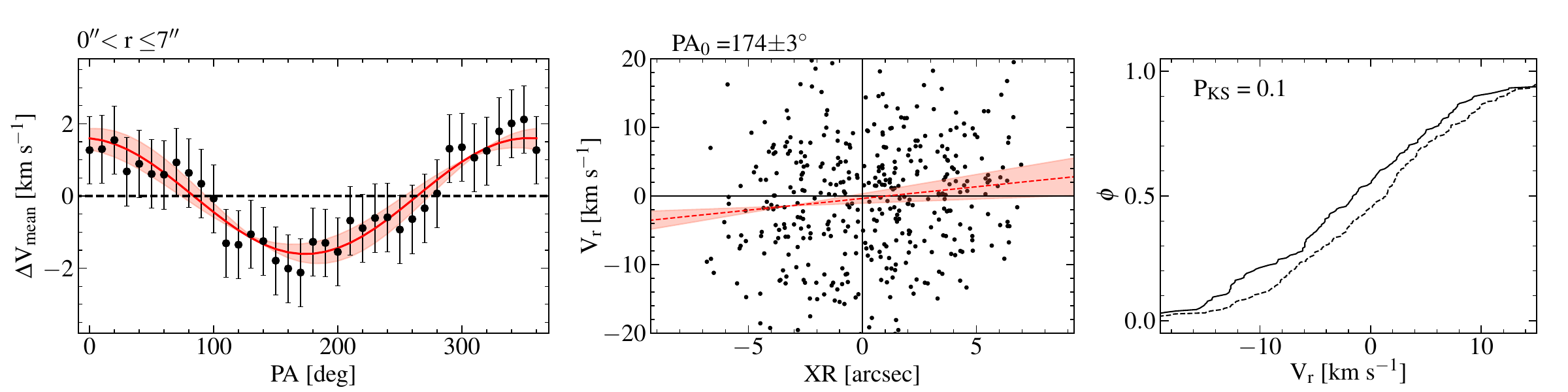}
\centering
\caption{Diagnostic diagrams of the rotation signal detected in M75,
  within 7\arcsec \ from the cluster center. {\it Left panel:}
  difference between the mean RV on each side of an axis passing
  through the center as a function of the position angle (PA) of the
  axis itself.  The red line is the sine function that best fits the
  observed pattern, while the red shaded region marks the confidence
  level at 3$\sigma$.  {\it Central panel:} distribution of the RVs
  referred to $V_{\rm sys}$ ($V_r$), as a function of the projected
  distances from the rotation axis (XR) in arcseconds.  The value of
  PA$_0$ is labeled.  The red dashed line is the least square fit to
  the data, and the red shaded area represents the 1$\sigma$
  uncertainty of the linear fit.
  {\it Right panel:} cumulative $V_r$ distributions for the
  stars with XR$<0$ (solid line) and for those with XR$>0$ (dotted
  line). The Kolmogorov-Smirnov probability that the two sub-samples
  are extracted from the same parent distribution is labeled.}
\label{fig:vrot}
\end{figure*}

The obtained results are discussed in the following and shown in
Fig. \ref{fig:vrot}.  Following the adopted method, the RV sample is
split into two sub-samples by a line passing through the center of the
system, and with position angle (PA) varying anticlockwise from
$0\degr$ (North) to $180\degr$ (South), by steps of $10\degr$.
The left panel of Fig. \ref{fig:vrot} shows the difference between
the average velocity of the stars on each side of this line ($\Delta
V_{\rm mean}$), as a function of PA.  This draws an approximately
sinusoidal pattern, which is a signal expected in the case of ordered
rotation.  From the maximum/minimum of the best-fit sine function (red
solid line), we derived the rotation amplitude ($A_{\rm rot} =
0.8 \pm 0.3$ km s$^{-1}$) as the half of this absolute value
and the position angle of the rotation axis
(PA$_0 = 174 \pm 3\degr$).  In the middle panel of the same figure,
the black dots show the measured RVs as a function of the projected
distances from the rotation axis (XR), the red dashed line is the linear fit to the data performed by means of a least square algorithm. A rotating cluster should show a
highly asymmetric distribution along two diagonally opposite
quadrants.  Finally, the cumulative velocity distributions of the two
sub-samples of stars on each side of the rotation axis are shown in
the right panel. A large difference between the two curves corresponds
to a high probability of rotation.  Therefore, to quantify the
statistical significance of this difference, we have used three
estimators. We computed the $p$-value of the Kolmogorov-Smirnov (KS)
probability that the RV distributions of the two sub-samples are
extracted from the same parent family ($P_{\rm KS}$), the $t$-Student
probability that the two RV samples have different means ($P_{\rm
  Stud}$), and the significance level (in units of n-$\sigma$) that
the two means are different following a Maximum-Likelihood approach
(n-$\sigma_{\rm ML}$), obtaining $P_{\rm KS}=0.1$, $P_{\rm Stud}>95\%$
and n-$\sigma_{\rm ML}=2.9$, respectively.

These results indicate a mild rotation signal in the region between
$0\arcsec$ and $7\arcsec$, with a maximum amplitude of $\sim 1.0$ km
s$^{-1}$.  Although the statistical significance of this signal is not
high, it is important to note that our estimate of the position angle
of the rotation axis (PA$_0 = 174 \pm 3\degr$) is in good agreement
with that found in \citet[][PA$_0 = -15 \pm 30\degr$, which
corresponds to 165$\degr$ \ according to our definition of
PA]{koch+18} using a sample of 32 stars in the radial region $20\arcsec< r < 250\arcsec$.
Considering that the cluster regions explored in these
two studies are complementary, we can conclude that
there is a hint of ordered rotation in M75.

\subsection{Velocity dispersion profile}
\label{sec:vdp}
Given a sample of individual RV measurements, the dispersion of the RV
distributions observed at different radial distances from the center
provides the second velocity moment profile $\sigma_{II}(r)$ of the
cluster. This is related to the projected velocity dispersion profile
$\sigma_P(r)$ through the following expression: $\sigma^2_P(r) =
\sigma^2_{II}(r) - A^2_{\rm rot}(r)$.  As discussed in the previous
section, the amplitude of internal rotation (if any) is very small,
and we can therefore assume that the velocity dispersion is reasonably
approximated by the second velocity moment (i.e., hereafter $\sigma_P(r) =
\sigma_{II}(r)$).

\begin{figure*}[ht!]
\centering
\includegraphics[width=0.9\textwidth]{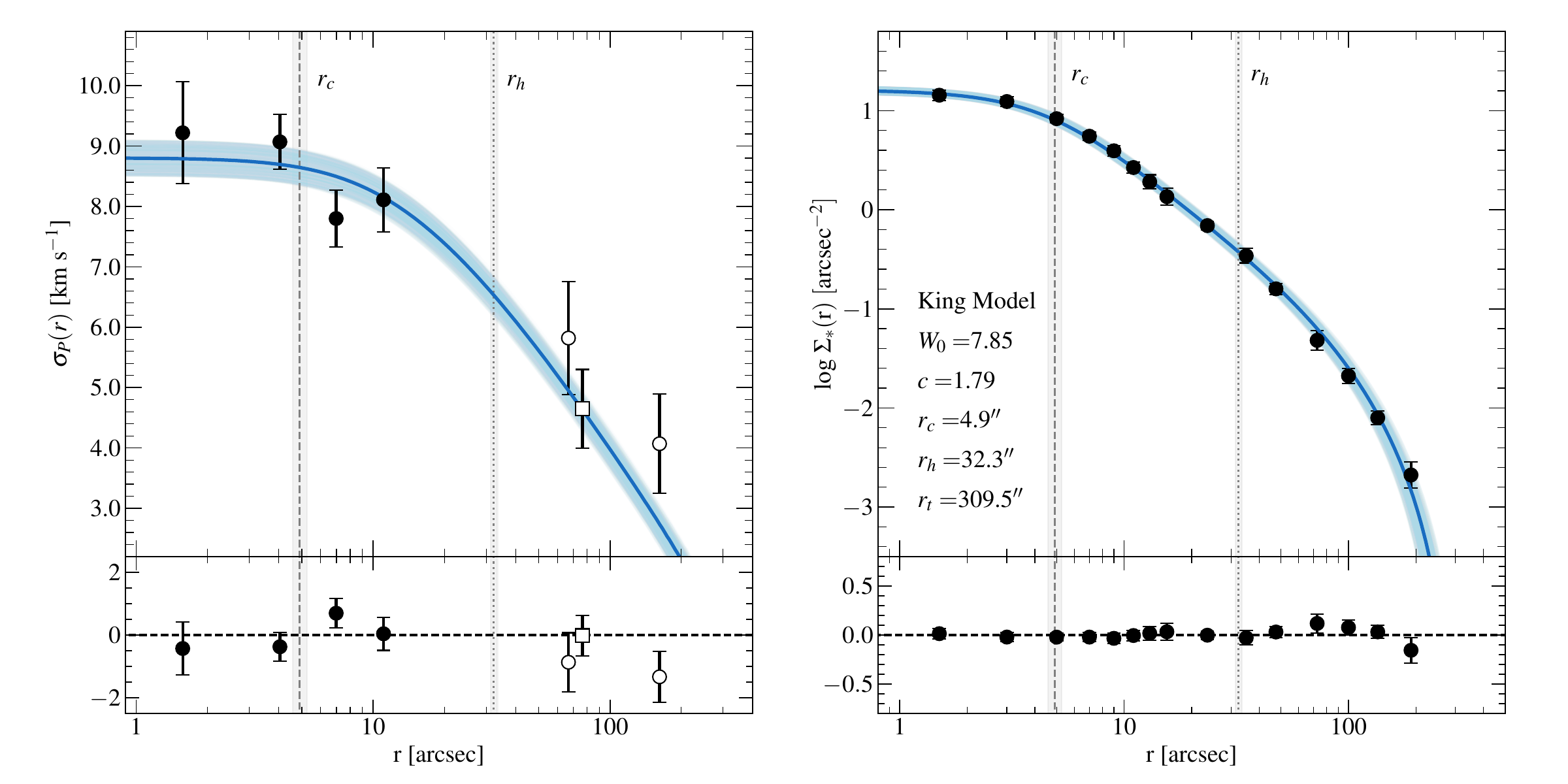}
\centering
\caption{\textit{Left panel}: Projected velocity dispersion profile of M75 obtained from the MUSE RV measures discussed in this work (black circles). The values of the line-of-sight velocity dispersion quoted in the online repository \protect\url{https://people.smp.uq.edu.au/HolgerBaumgardt/globular/}, and the one obtained from \textit{Gaia} proper motions \citep{vasiliev+21} rescaled to the distance estimated in Section \ref{sec:dist_age} (see Table \ref{tab_final}), are also plotted as empty circles and empty square, respectively.  The blue line corresponds to the best-fit King model derived from the simultaneous fit to this profile and the density profile shown in the right panel.  The shaded regions show the $1\sigma$ uncertainty on the best-fit. The dashed and dotted vertical lines mark the derived core radius ($r_c$) and half-mass radius ($r_h$), respectively.  The bottom panel shows the residuals between the King model and the observations. \textit{Right panel}: Projected density profile from resolved star counts obtained by \citet[][black circles]{contreras+12}. The meaning of the lines, shaded region, and bottom panel is as in the left panel.  The values of the best-fit central dimensionless potential ($W_0$), concentration parameter ($c$), $r_c$, $r_h$, and tidal radius ($r_t$) are labeled in the panel.}
\label{fig:vdp}
\end{figure*}

To determine the velocity dispersion profile of M75 we divided the
sample of bona fide RV measures in four concentric radial bins at
increasing distance from the center, each one including more than 50 stars. 
We then applied a 3$\sigma-$clipping algorithm to remove the
outliers, and we determined the dispersion of the remaining RV
measures through a Maximum-Likelihood approach, by maximizing the
following likelihood:
\begin{equation}
  \mathrm{ln}\mathcal{L}=-\frac{1}{2}\sum_{i}^{}\Big[\frac{V_{r,i}^2}{(\sigma^2+\epsilon_{v,i}^2)}
    + \mathrm{ln}(\sigma^2+\epsilon_{v,i}^2)\Big]
\end{equation}
where $V_{r,i} = RV_i -V_{\rm sys}$ is the RV of the $i^{th}$ star in
the considered bin with respect to the cluster systemic velocity,
$\epsilon_{v,i}$ is its corresponding error, $\sigma$ is the velocity
dispersion in the bin, and the summation is made over all the stars
included in the bin. We implemented the likelihood using a Markov
Chain Monte Carlo (MCMC) approach, based on the emcee algorithm
\citep{Foreman+13}.  From the posterior probability distribution
function (PDF) obtained for the velocity dispersion in each bin, we
derive the best-fit values of the dispersion as the PDF median, and
its uncertainty from the $68\%$ confidence interval.  The velocity
dispersion profile thus determined in the core of M75 is shown in the
left panel of Fig. \ref{fig:vdp} (black circles) and listed in Table
\ref{tab:vdp}.  The profile is roughly constant in the very center and
starts to decrease at about $6\arcsec$ from the center.

\begin{table}
\caption{Velocity dispersion profile of M75.}
\centering
\setlength{\tabcolsep}{7pt} 
\renewcommand{\arraystretch}{1.2} 
\begin{tabular}{crrccc}
\hline\hline
 $r_i$ [$\arcsec$] & $r_e$ [$\arcsec$] & $r_m$ [$\arcsec$] & $N$ & $\sigma_P$ km s$^{-1}$& $\epsilon_{\sigma_P}$ km s$^{-1}$ \\
\hline
0.01  &  2.40  &  1.57 & 69  & 9.2 &  0.9  \\
2.40  &  5.70 &  4.04  & 240 & 9.1 &  0.5  \\
5.70  &  8.50 &  6.98  & 171 & 7.8 &  0.5  \\
8.50  & 16.00 & 11.06  & 146 & 8.1 &  0.6 \\
\hline
\end{tabular}
\tablefoot{The table lists: the internal, external and mean radii of each
adopted radial bin ($r_i$, $r_e$, and $r_m$, respectively), the
number of stars in the bin ($N$), the projected velocity dispersion ($\sigma_P$) and its uncertainty in the bin ($\epsilon_{\sigma_P}$).}
\label{tab:vdp}
\end{table}

\section{Distance, reddening and age estimate}
\label{sec:dist_age}
The high quality of the available HST photometric data set offers the
possibility of updated estimates of the cluster distance, reddening
and age. In fact, the positions of various evolutionary sequences in
the CMD of an old stellar population provide useful references to
constrain these parameters through the comparison with theoretical
models \citep[see examples in][]{saracino+19, cadelano+19,
 cadelano+20a, cadelano+20b, deras+23, massari+23}.
In particular, the vertical portion of the CMD at the base of the RGB, and the horizontal portion of the horizontal branch (HB) are sensitive, respectively, to the color excess and the cluster distance, independently of the (old) age of the population, while the main
sequence turnoff and the sub-giant branch regions are the most age-sensitive evolutionary sequences.

To determine first-guess estimates of color excess and distance,
we extracted from the BaSTI-IAC database \citep{pietrinferni+2021} a 12 Gyr old isochrone with appropriate metallicity ([Fe/H]$=-1.3$ dex; \citealt{carretta+09}) 
standard helium mass fraction ($Y$ = 0.25), and [$\alpha$/Fe]$= +0.4$ (which is the typical value for Galactic GCs at this metallicity).
We searched for the shifts in color and magnitude
necessary to superpose the model to the samples of stars selected at
the base of the RGB in the magnitude range $19< m_{\rm F555W}<20$ and
in the HB portion at $0.5< (m_{\rm F438W}-m_{\rm F555W}) <0.8$.
A reasonable reproduction of the observed loci is obtained by adopting 
a color excess $E(B-V) = 0.16$ and a distance modulus $(m-M)_0 = 16.7$.
Starting from these first-guess values, we then determined the best-fit cluster age, distance, and reddening through the isochrone fitting technique, 
following a Bayesian procedure similar to that used in many previous works \citep[see, e.g.,][]{saracino+19, cadelano+19, cadelano+20b, deras+23, massari+23}. 
This approach consists in comparing the observed CMD
with a set of isochrones, simultaneously exploring suitable grids for the age, distance modulus and color excess.
We extracted a set of isochrones from the BaSTI-IAC repository \citep{pietrinferni+2021} 
computed assuming the same chemical composition quoted above, and ages ranging between 9 Gyr and 15 Gyr, in steps of 0.2 Gyr.
The comparison between the observed CMD and the isochrones has been performed by adopting an MCMC technique, assuming a Gaussian likelihood function \citep[see equations 2 and 3 in][]{cadelano+20b}.
To properly constrain the level of the HB, 
we also included a term in the likelihood function to minimize the distance in magnitude between the zero-age horizontal branch of the models and the level of the red portion of the observed HB.
We  used the emcee code \citep{Foreman+13} to sample the posterior probability distribution in the parameter space.
A flat prior has been assumed for the explored age range.
To convert the isochrone absolute magnitudes to the observational plane,
we adopted values of the color excess and distance modulus following Gaussian prior distributions peaked at the first-guess values quoted above.
To improve the definition of the main sequence turnoff and sub-giant
branch evolutionary sequences in the CMD (in the magnitude range $19.5<
m_{\rm F555W}<21.5$), we considered only stars at distances larger
than $25\arcsec$ from the cluster center.

\begin{figure*}[ht!]
    \centering
    \includegraphics[scale=0.35]{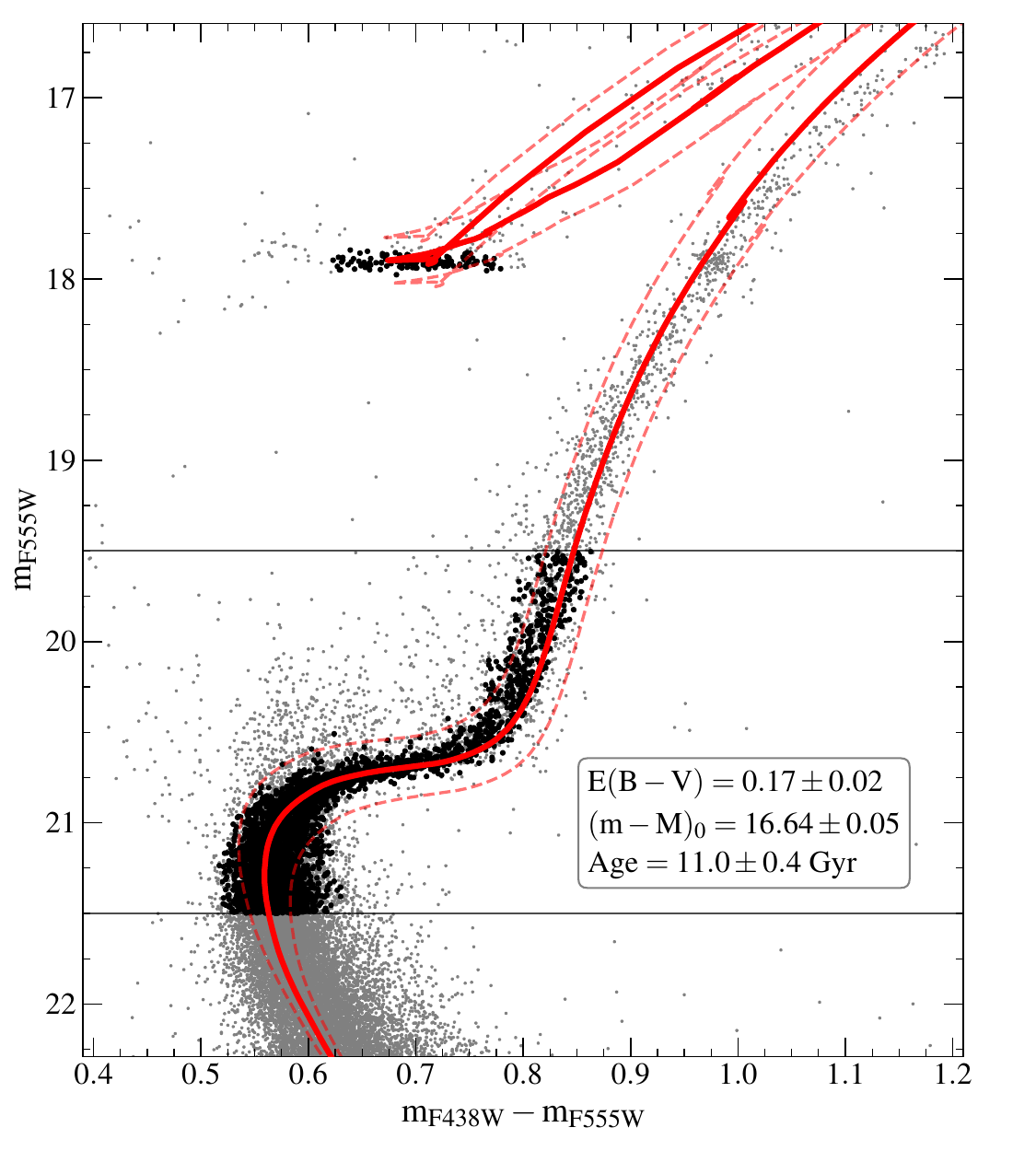}
    \includegraphics[scale=0.3]{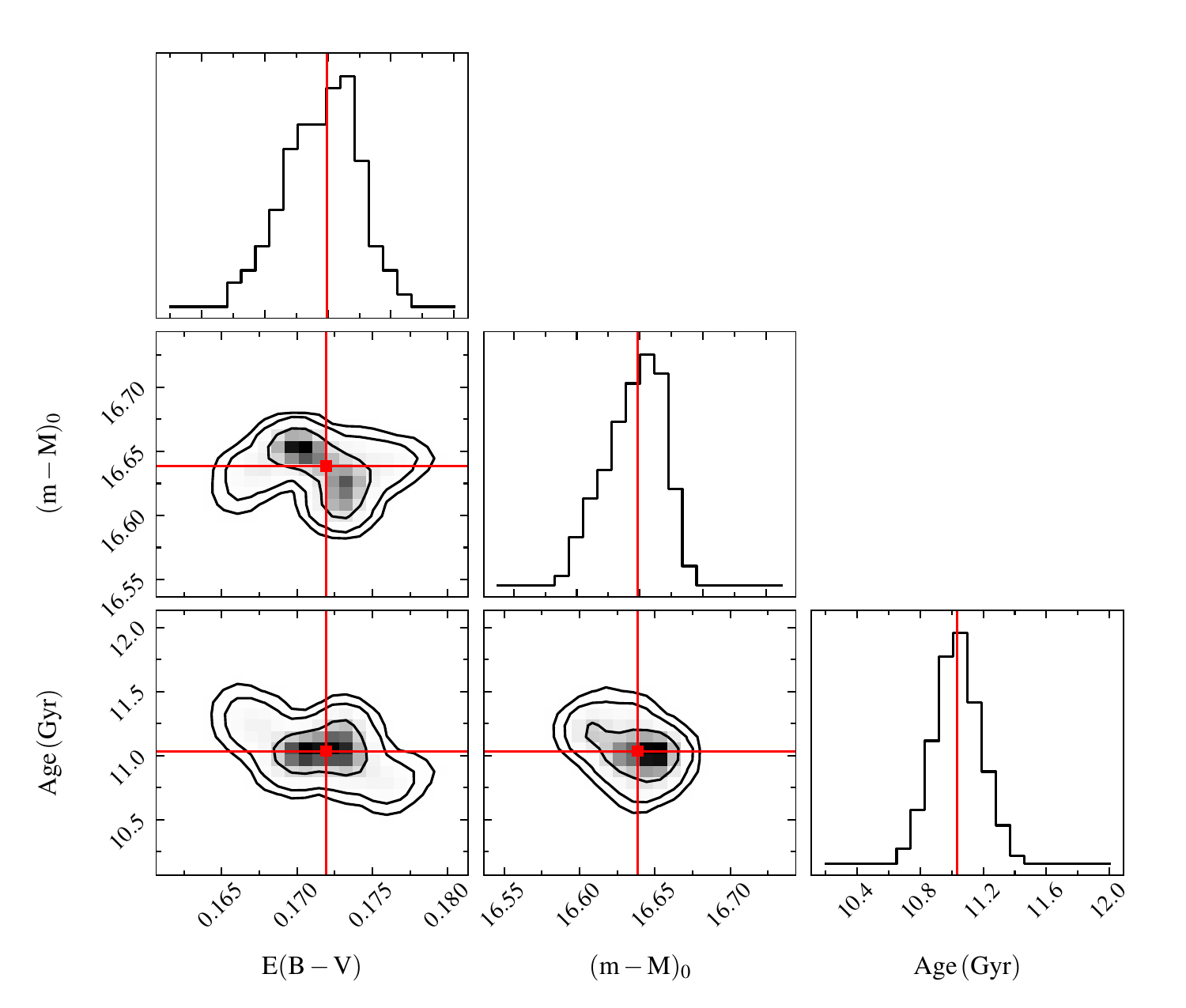}    
    \caption{\textit{Left panel}: CMD of M75 (gray dots) with the best-ﬁt BaSTI isochrone
    computed with [Fe/H]$=-1.3$ dex plotted as a red solid line. 
    The dashed red isochrones represent the uncertainties on the parameters. The black circles mark the stars used for the isochrone fitting procedure. The resulting best-fit values of color excess, distance modulus and age are labeled.
    \textit{Right panel}: corner plots showing the one- and two-dimensional projections of the posterior probability distributions
for all the parameters derived from the MCMC method. 
 The contours correspond to the 68\%, 95\%, and 99\% confidence levels.}
\label{fig:age}
\end{figure*}

The best-ﬁt values correspond to the 50th percentile of the posterior probability distribution of each parameter.
Since we have constrained the HB level and fixed the metallicity of the models, the formal $1\sigma$ uncertainties obtained from the
posterior probability distributions of the MCMC are underestimated. Hence, we have included the uncertainty on the HB level in the propagation of the errors and adopted the conservative $1\sigma$ uncertainties listed below.
The best-fit values of the derived quantities are
 $E(B-V) = 0.17 \pm 0.02$, $(m-M)_0 = 16.64 \pm 0.05$
(corresponding to a distance of 21.3 $\pm \ 0.5$ kpc), and an 
age of 11.0 $\pm \ 0.4$ Gyr (see also Table \ref{tab_final}). 
The resulting best-ﬁt isochrone is shown in Fig. \ref{fig:age}.
Assuming a larger metallicity 
([Fe/H]$= -1.2$ dex; see \citealt{Kacharov+13}), the isochrone fit slightly worsens, but it still provides values of the three parameters that are well consistent with the best-fit ones within the errors.
While the estimated reddening is
fully consistent with that reported in the literature \citep{harris+96},
the distance of the cluster is slightly larger than the previous values
(20.9 kpc in \citealp{harris+96}, 20.52 kpc in \citealp{baum_vasil+21}).

\section{Discussion and Conclusions}
\label{sec:discussion}
In this paper we presented a new result of the MIKiS survey: the first
kinematic exploration of the central regions of M75 (NGC 6864), a GC
orbiting the outer halo of the Milky Way. By using a set of
AO-corrected MUSE/NFM observations, we obtained a sample of $\sim
1900$ RV measurements of individual stars located within $16\arcsec$
($\sim 3r_c$)
from the cluster center.  From a sub-sample of 630 well measured stars
(i.e., with MUSE spectra having S/N$>20$ and with RV uncertainties
smaller than 5 km s$^{-1}$), we determined the systemic velocity of
the cluster ($V_{\rm \rm sys} = -189.5 \pm 0.3$ km s$^{-1}$) and the
innermost portion ($r<16\arcsec$) of its velocity dispersion profile.
By taking advantage of the high photometric quality of the HST/WFC3
catalog, we also determined updated estimates of the cluster
distance, reddening and age (see Table \ref{tab_final}).

By using a set of single-mass, spherical, isotropic and non-rotating
King models \citep{king+96},\footnote{The King models \citep{king+96}
are a single-parameter family of dynamical models, meaning that their
shape is univocally determined by the dimensionless parameter $W_0$,
which is proportional to the gravitational potential at the center of
the system, or alternatively, by the concentration parameter
$c\equiv\mathrm{log}(r_t/r_0)$, where $r_t$ and $r_0$ are the tidal
and the King radii of the model, respectively. Other characteristic
parameters of the cluster structure are the core radius $r_c$ and the
half-mass radius $r_h$.} We simultaneously fitted the velocity
dispersion profile obtained by combining the one derived in this work with literature values (see left panel of Fig. \ref{fig:vdp}), and the projected density profile
obtained from the resolved star counts in \citet{contreras+12}.  The
approximation of single-mass and non-rotating models is legitimate by
the fact that both profiles have been obtained from samples of stars
with approximately the same mass (main sequence turnoff and giant
stars), and that the rotation signal is negligible in this system (see
Section \ref{sec:rot}).

The best-fit solution has been determined by using an MCMC method
through the emcee algorithm \citep{Foreman+13}. We assumed a 
$\chi^2$ likelihood and uniform priors on the parameters of the fit 
(i.e., the King concentration parameter $c$, the core radius $r_c$,
the value of the central density, and the central velocity dispersion $\sigma_0$).
For each parameter, the 50-th percentile of the PDF has been adopted as best-fit value,
while the 16-th and 84-th percentiles have been used to determine the
$1\sigma$ uncertainty.  The fitting procedure provides a best-fit King
model characterized by $W_0 =7.85\pm0.1 $, concentration
$c=1.79\pm 0.03$, core radius $r_c=4.9{\arcsec}_{-0.3}^{+0.4}$,
half-mass radius $r_h=32.3{\arcsec}_{-1.0}^{+1.2}$, tidal radius
$r_t=309.5{\arcsec}_{-10.7}^{+11.9}$, and a central velocity dispersion
$\sigma_0 = 8.8 \pm 0.3 $ km s$^{-1}$. 
Figure \ref{fig:vdp} shows the resulting best-fit King model (blue
solid lines) overplotted to the observed velocity dispersion and
density profiles (left and right panels, respectively).  The bottom
panels show the residuals between the model and the observations.  The
best-fit values and the uncertainties of each parameter are listed in
Table \ref{tab_final}.\\
The value of the central velocity dispersion estimated in this work is
significantly smaller than those quoted in \citet[][$\sigma_0 =
  14.1_{-2.0}^{+2.4}$ km s$^{-1}$]{koch+18}, and in \citet[][$\sigma_0
  = 11.8$ km s$^{-1}$, which has been updated to 12.3 km s$^{-1}$ in
  the online version]{Baumgardt+18}.  However, the literature values
are both extrapolated from the fit to observational data that are
confined to the cluster outer regions, while, for the first time in
this work, the central portion of the velocity dispersion profile is
directly constrained by observations.  In particular, \citet{koch+18}
fitted a Plummer model (see their eq. 3) to a sample of just 32 stars
located between $\sim$20\arcsec \ and 250\arcsec \ from the center, while \citet{Baumgardt+18} used N-body simulations and a
velocity dispersion profile with the innermost radial bin located at
$\sim 70\arcsec$ from the center (see empty circles in
Fig. \ref{fig:vdp}).
It is also worth keeping in mind that our analysis is mainly based on
RGB stars that are slightly more massive ($\sim 0.8 M_\odot$) than the
average cluster stars ($\sim 0.3 M_\odot$), while the central velocity
dispersion estimated by \citet{Baumgardt+18} from N-body simulations
is weighted by stellar masses.  Taking into account the effects of
energy equipartition and mass segregation, this could partially
justify the smaller value estimated in this work.  Indeed, the
analyses conducted so far clearly suggest that M75 is a highly
dynamically evolved stellar system. The central relaxation time
\citep[$\mathrm{log} t_{rc}/$yr = 8.0][]{lanzoni+16} compared to the
cluster age (11 Gyr; see Section \ref{sec:dist_age}) indicates that the
system has undergone several relaxations so far. In addition, the
large value of the BSS segregation level (the $A_{rh}^+$ parameter)
empirically confirms the high level of dynamical evolution and central
mass segregation in this cluster \citep[see,][]{lanzoni+16,
  ferraro+23a}. Hence, RGB stars are more centrally segregated than
the average and their central velocity dispersion is lower than the
mass-weighted value.

As motivated above, M75 can be well approximated by a single-mass,
spherical, isotropic, and non-rotating King model
(\citealt{king+96}). Under this assumption, the total mass of the
system can be estimated from the value of $\sigma_0$ following
equation (3) in \citet[][see also
  \citealt{Richstone+86}]{Majewski+03}: $M = 1.66 r_c \mu/\beta$,
where $\mu$ is a polynomial function of the King concentration
parameter (see eq. 8 in \citealt{Djorgovski+93}), and $\beta =
1/\sigma_0^2$.  The resulting total mass is $M = 2.5\pm 0.2 \times
10^5 M_\odot$, where
the uncertainty has been estimated through Monte Carlo simulations
\citep[see][]{leanza+22, pallanca+23}.  This value is a factor of two
smaller than the value derived in \citet[][$5.86 \pm 1.24 \times 10
  ^5 M_\odot$]{Baumgardt+18}, mainly due to the significant difference
in the values of $\sigma_0$, and also because of the different
assumptions (e.g., from Section \ref{sec:dist_age}, 
we adopted a distance of
21.3 kpc, while \citealt{Baumgardt+18} use 20.5 kpc) and the different
methods used in the two works.

Finally, using the selected sample of MUSE RVs, we investigated the
systemic rotation in the innermost region of the cluster
($r<7\arcsec$, corresponding to $r<1.4 r_c$).  We found just a weak hint of rotation ($A_{\rm
  rot}\sim 1$ km s$^{-1}$) with a position angle of the rotation axis
PA$_0 = 174 \pm 3 \degr$. 
It is intriguing to note that the value of
PA$_0$ obtained in the present paper is fully consistent with that of the rotation signal detected
by \citet{koch+18} in the outer regions.  Thus, in spite of the
limitation of the two samples and the low statistical significance of both detections, there is now a stronger hint of internal rotation in M75 that deserves further investigation.

\begin{table}
\caption{Parameters of M75 determined in this work.}
\centering
\small
\setlength{\tabcolsep}{11pt} 
\renewcommand{\arraystretch}{1.4} 
\begin{tabular}{lc}
\hline\hline
Parameter & Estimated Value \\
\hline
    Color excess & $E(B-V) = 0.17\pm0.02$ \\
    Distance modulus &  $(m-M)_0 = 16.64\pm0.05$ \\
    Cluster distance & $d=21.3 \pm0.5$ kpc \\
    Age & $ 11.0\pm0.4$ Gyr \\
    Dimensionless central potential  & $W_0 = 7.85\pm 0.1$   \\
    Concentration parameter & $c=1.79\pm 0.03$  \\
    Core radius & $r_c=4.90{\arcsec}^{+0.4}_{-0.3}$  \\
    Half-mass radius & $r_{\rm h}=32.3{\arcsec}_{-1.0}^{+1.2}$  \\
    Tidal radius & $r_t=309.5{\arcsec}_{-10.7}^{+11.9}$  \\
    Systemic velocity & $V_{\rm sys} = -189.5 \pm 0.3$ km s$^{-1}$  \\
    Central velocity dispersion & $\sigma_0 = 8.8 \pm 0.3$ km s$^{-1}$   \\
    Total mass & $M = 2.5 \pm0.2 \times 10^5 M_\odot$  \\
\hline
\end{tabular}
\label{tab_final}
\end{table}

\begin{acknowledgements}
This work is part of the project {\it Cosmic-Lab} at the Physics and
Astronomy Department "A. Righi" of the Bologna University
(http://www.cosmic-lab.eu/ Cosmic-Lab/Home.html).  
\end{acknowledgements}

%
%


\end{document}